\documentclass[runningheads]{llncs}
\usepackage[T1]{fontenc}
\usepackage{caption}
\usepackage{graphicx}
\usepackage{xcolor}
\usepackage[hidelinks]{hyperref}
\usepackage{color}

\begin{document}
\title{Random boundaries: quantifying segmentation uncertainty in solutions to boundary-value problems}
\titlerunning{Random boundaries}
%
\author{S. G. Gralton\textsuperscript{*}\inst{1}\orcidID{0000-0001-7259-1549} \and
F. Alkhatib\inst{1}\orcidID{0000-0003-3367-0353} \and
B. Zwick\inst{1}\orcidID{0000-0003-0184-1082} \and
G. Bourantas\inst{2}\orcidID{0000-0001-9503-4614}\and
A. Wittek\inst{1}\orcidID{0000-0001-9879-8361}\and
K. Miller\inst{1,3}\orcidID{0000-0002-6577-2082}}
\authorrunning{S. G. Gralton et al.}
%
\institute{Intelligent Systems for Medicine Laboratory, University of Western Australia, Australia \and
University of Patras, Greece \and
Harvard Medical School, MA, USA }
\maketitle              
\begin{abstract}
Engineering simulations using boundary-value partial differential equations often implicitly assume that the uncertainty in the location of the boundary has a negligible impact on the output of the simulation. In this work, we develop a novel method for describing the geometric uncertainty in image-derived models and use a naive method for subsequently quantifying a simulation's sensitivity to that uncertainty. A Gaussian random field is constructed to represent the space of possible geometries, based on image-derived quantities such as pixel size, which can then be used to probe the simulation's output space. The algorithm is demonstrated with examples from biomechanics where patient-specific geometries are often segmented from low-resolution, three-dimensional images. The results show the method's wide applicability with examples using linear elasticity and fluid dynamics. We show that important biomechanical outputs of these example simulations, maximum principal stress and wall shear stress, can be highly sensitive to realistic uncertainties in geometry.

\end{abstract}
\section{Introduction}
Boundary-value problems (BVP) are a versatile and widely used mathematical model for simulating engineering problems from a diverse array of fields including manufacturing, aeronautics and medicine. In general, a BVP is composed of a set of differential equations, boundary conditions, constitutive properties, an applied load and a domain on which the problem geometry is defined. These features make BVPs a highly versatile mathematical model which can simulate a large variety of different physical processes.

Most practitioners using these types of simulations are conscious that solutions to BVPs can be highly sensitive to poorly specified loads and constitutive properties. The stochastic finite element method (SFEM) literature has devoted a large body of research to developing methods for quantifying the uncertainties stemming from the parameterisation of these variables \cite{Arregui2016,Ghanem1991,Stefanou2009}.

In contrast, the uncertainty quantification community has paid relatively little attention to understanding these simulation's sensitivity towards a poorly described geometry. As the boundary forms the basis on which all other inputs are defined, it is reasonable to hypothesise that the boundary's location could have a significant effect on the simulation's output. In general, most engineering simulations model scenarios where an object's boundaries are measurable quantities that have been produced through manufacturing processes with toleranced geometric deviations. In such a scenario, these tolerances are usually sufficiently low that the inaccuracies in other simulation parameters (constitutive properties, etc) will have significantly more impact on the results and thus, are more deserving of uncertainty quantification efforts.

However, in cases where BVPs are applied on less rigorously constrained geometries, we should consider the effects of geometric inaccuracy to be as important as that of any other simulation parameter. Quantifying these inaccuracies and their effect is something that to date has only been explored to a limited extent.

This paper develops an easily applicable method for quantifying uncertainty from geometric data with finite resolution such as an image. This is particularly relevant in fields such as patient-specific bioengineering which relies on geometry extracted from low-resolution images to calculate important clinical quantities. The input geometry is described by a Gaussian random field and the original segmented geometry. With the geometric uncertainty thus parameterised a direct Monte Carlo algorithm or any other sampling methodology can be used to probe a simulation's output space. The main contribution of this paper is in the formulation of a realistic and meaningful representation of the uncertainty in image-derived geometries.

\subsection{Bioengineering context}
The need for a robust method to quantify geometric uncertainty is particularly apparent in the modern bioengineering literature. Bioengineering simulations often use three-dimensional images of a patient's anatomy, such as those produced by magnetic resonance imaging (MRI) or computed tomography (CT), as the basis for geometric models. Patient-specific simulations are designed to use the unique and varied anatomy of each individual to generate relevant clinical data that can aid decision making. Current standards for generating patient-specific models are based on segmenting the anatomical structure of interest from medical images. Accurate and automated segmentation processes are still an ongoing area of research but even if they were perfect, the resolution, blur and noise of most medical images places an incontrovertible limit on the segmentation's accuracy \cite{Warfield2002}.Multiple studies have investigated and acknowledged large variations in segmented geometry due to differences in human input, algorithms and images \cite{Hoyte2009,Hoyte2011,Velasco-Annis2018}.

\begin{figure}
  \captionsetup{width=\linewidth}
  \begin{center}
  \includegraphics[width=0.5\linewidth]{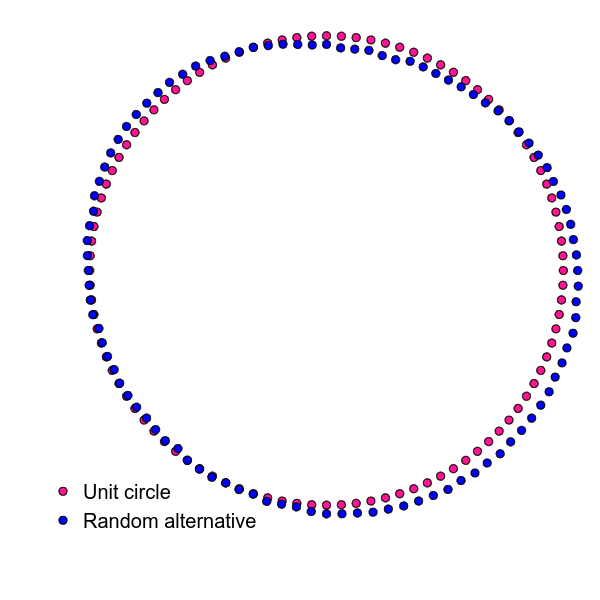}
  \end{center}
  \caption{A unit circle and a randomly generated alternative geometry. A squared-exponential covariance kernel (Equation \ref{kernel_eq}) was used with $\sigma=0.05$ and $l=0.8$.}
  \label{fig:boundary}
\end{figure}

Due to the known challenges of imaged-based segmentation many bioengineering researchers have attempted to demonstrate their simulations' insensitivity to an uncertain geometry. An early attempt to quantify this emerged in 1994 with \cite{Chinchalkar1994} but has not advanced significantly since. In general, a Monte Carlo approach has been used with a variety of different methods for generating the input space of geometries. Some studies have applied various affine transformations of the original geometry to generate their samples \cite{Bruning2018,Mayeur2016,Sankaran2015,Taddei2006}. Others have generated a suite of possible geometries by segmenting the same images multiple times using different analysts or techniques \cite{Bertolini2022,Joldes2017,Little2015,Nikhilesh2021,Toma2021}. Further works have reduced the geometry to key dimensions such as maximum diameter, assigned a distribution to those variables and then randomly sampled them and constructed models to fit \cite{Arsene2013,Celi2013,Donaldson2014,Niemeyer2012}. Interestingly, many of these studies found significant sensitivity to geometric uncertainty in their results reinforcing the need for these studies. However, two major flaws exist throughout all existing methods; they either require significant human input or dramatically simplify the structure of the spatial uncertainty in the data source. Both of these flaws are problematic but in particular, the simplifications of the uncertainty distribution ignore the possibility for intra-sample randomness in any given segmentation, something we aim to address.

In contrast, this paper's proposed method uses the image resolution to inform the magnitude of any potential deviations in the boundary and the distance over which these deviations may exist. The image resolution, or some small multiple of it, is a good basis for these values as any details below the pixel size are necessarily obscured. With the magnitude and length scale of the spatial uncertainty quantified, a library of representative, alternative geometries can be generated. An example of a random alternative to a unit circle generated using the proposed method has been included in Figure \ref{fig:boundary}. Using this simple parameterisation infinitely many geometries can be generated, all of which reflect the uncertainty in the original segmentation. This allows us to model geometric uncertainty in a way that strongly reflects the actual structure of the uncertainty in a repeatable manner.

This paper outlines a new, simple method for quantifying geometric uncertainty. The method is then demonstrated by calculating uncertainties in wall shear stress in a nominally perfect cylinder and maximum principal stress in a patient-specific model of an abdominal aortic aneurysm segmented from a 3D computed tomography (CT) image.

\section{Method}\label{method}

\begin{figure}
  \captionsetup{width=\linewidth}
  \begin{center}
  \includegraphics[width=0.85\linewidth]{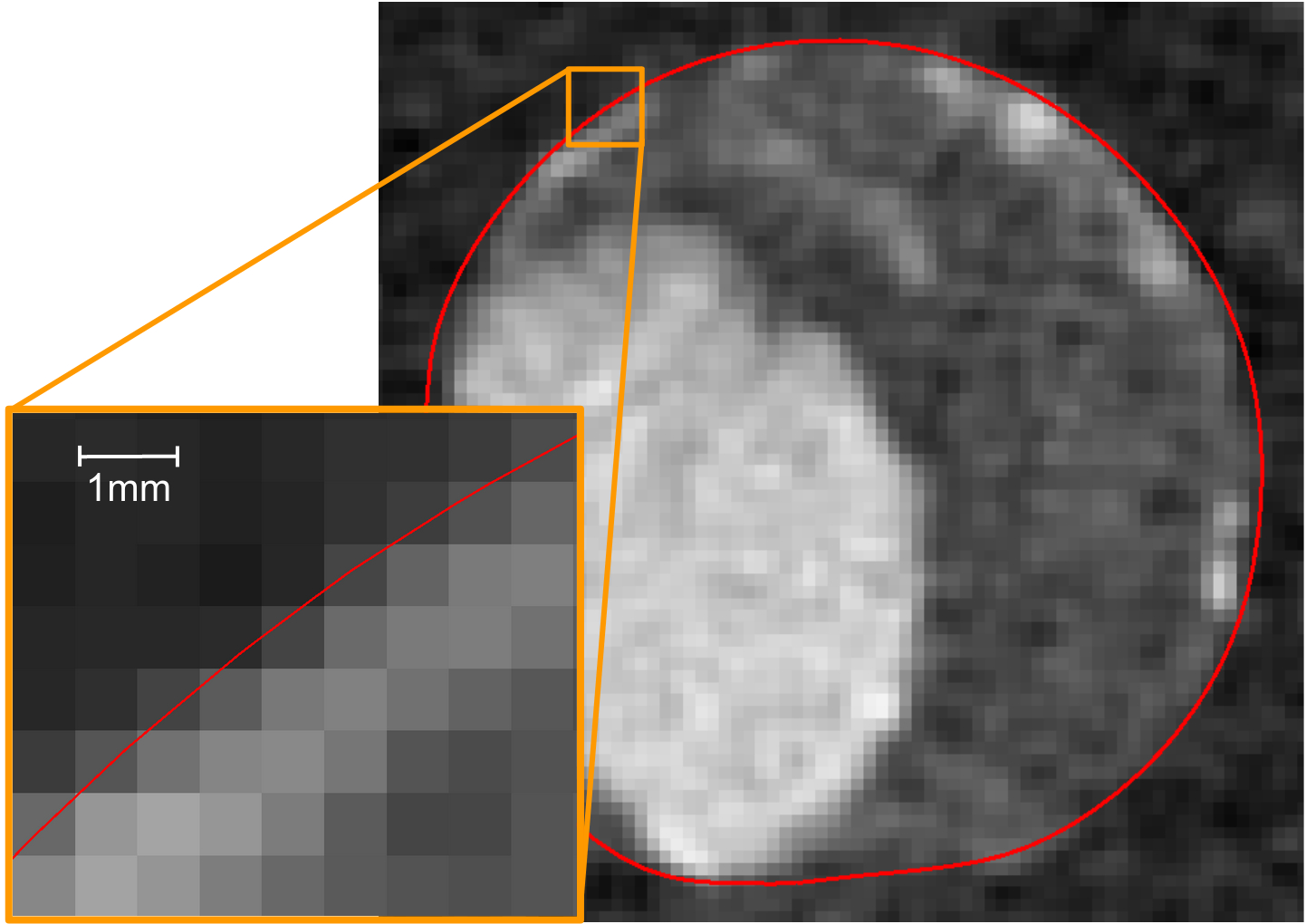}
  \end{center}
  \caption{An extract from an computed-tomography image with the segmented boundary of the abdominal aortic aneurysm marked in red. The zoomed section is of particularly clear part of the image where the wall thickness can be clearly seen. The combination of the image's noise and resolution limits the accuracy of the outer boundary segmentation to within one to two voxels. }
  \label{fig:boundary_example}
\end{figure}

Monte Carlo (MC) algorithms offer a simple but computationally expensive way to probe the output space of a simulation through sampling of the input space. Implementing an effective MC simulation to find geometric uncertainty requires generating a library of geometries which reflect the uncertainty present in the geometric data. To generate a meaningful uncertainty on the output, this library of geometries should sample the space of possible geometries that could reasonably be extracted from the dataset used to generate the initial geometry. These geometries are compatible with more efficient methods for sampling the output space such as polynomial chaos expansion but a direct MC algorithm has been used for illustrative purposes and to build incrementally on previous attempts at quantifying this uncertainty.

The space of possible geometries in the proposed methodology takes the format of a Gaussian random field over the geometric domain which indicates at any point on the model the magnitude with which to move the boundary in a direction normal to the local surface. This structure reflects an assumption that the segmented boundary at any point on the model is a random draw from a normal distribution centred on the true physical boundary with a width that corresponds closely to the resolution of the image. Obviously the true boundary location is unavailable so the segmented boundary is warped with a Gaussian random field in order to sample the space within which the true boundary could exist. The use of a Gaussian random field allows the geometric uncertainty to be explicitly modelled with a prescribable amplitude and smoothness; two variables that can be easily extracted from a qualitative image assessment.

A spatial random field is a convenient and realistic way to describe geometric uncertainty only if it can be assumed that the segmented boundary is correct within the limitations of image resolution. This means that random boundaries will not account for gross segmentation errors and can only model an uncertain boundary location due to insufficient image resolution or similar statistical errors. If this assumption can be made for a given model then the geometric uncertainty for that model can be parameterised by the maximum amplitude of potential deviations from the segmented boundary and some smoothness parameter to enforce the spatial frequency of those deviations.

\subsection{Geometric uncertainty parameterisation}
Random fields describe a stochastic process over a domain. At any point in a given domain, a draw can be made from a random field to generate a random scalar that follows the global properties of the entire field. A Gaussian random field has two important global properties, namely a prescribed mean and variance over the entire domain. Additionally, a given field can be structured in such a way that two neighbouring points are spatially covariant, i.e. if one point has sampled a high value from the random field it is more likely that the adjacent locations will also have a high value.

The uncertainty in the boundary location of a poorly-resolved image can be described with a Gaussian random field over the entire model domain. We will construct this random field and a draw from it at some location in the domain, $x_i$, in the following manner:
\begin{equation} \label{field_eq}
    p(x_i) \sim \mathcal{N} \left( \mu, C(X, X') \right)
\end{equation}

where $\mu$ is the mean deviation over the domain and $C(X,X')$ is the covariance between each each spatial point, $x_i$, in the set of all points describing the surface, $X$.

The first variable, $\mu$, implies in this context the mean deviation of the alternative geometry space to the segmented boundary. In a non-biased example it is reasonable to use $\mu = 0$mm but if a global bias exists, such as from thresholding or a smoothing operation, this should be captured in by the mean deviation.

The key to generating random geometries lies in selecting a relevant and physically informed covariance structure for the random field, $C(X,X')$. The first component of this is to select a covariance kernel. Selecting a covariance kernel can be a problematic issue and in much of the stochastic FEM literature the choice is left to the user, however, the choice of kernel has been found to have limited effect on the output of random boundaries. As long as the kernel is smooth, similar results were observed regardless of the covariance kernel used as shown in Section \ref{lin_elastic}. A squared-exponential kernel is used below as an illustrative example:
\begin{equation} \label{kernel_eq}
    C(x_i, x_j) = \sigma ^2 \exp{ \left( - \frac{(x_i - x_j)^2}{l^2} \right) }
\end{equation}
where $C(x_i, x_j)$ is the covariance between two points, $x_i$ and $x_j$, $\sigma$ is a scaling factor and $l$ is the length scale. For most kernels, the choice of scaling and length parameters are the dominant factors that can be adjusted to represent a given dataset's scale and structure of uncertainty. Section \ref{num_examples} gives guidelines for how this has been approached for each example.

The distance between two points, $x_i$ and $x_j$, can be calculated using any norm if the random field will be generated using eigenvalue decomposition. It would ideally be the geodesic distance between suitably spaced nodes on the surface but in practise this becomes difficult as models get larger and eigenvalue decomposition gets significantly more expensive. The numerical examples in Section \ref{num_examples} use the Euclidean distance between points as a convenient surrogate for the geodesic distance.

The length scale, $l$, and magnitude, $\sigma$, of the kernels are the most important parameters when describing the covariance structure and some analogue of them exists regardless of which covariance kernel has been selected. A qualitative analysis of the image describing the object of interest should result in a reasonable quantification for these two important variables. Figure \ref{fig:boundary_example} is an example of an uncertain boundary described by an image. In this example, the gradient across the segmented boundary implies that the true boundary could exist in any of the two pixels either side of the segmented boundary. Under the assumption that the segmented boundary is correct in a mean sense across the whole of the image, the standard deviation for the boundary position normal to the boundary could be thought of as having a 95\% probability of being within 2 pixels either side of the segmented boundary and therefore, $\sigma \approx 1mm$ or one pixel width.

The length scale of the covariance kernel, $l$, is harder to parameterise but it represents something akin to a smoothness parameter. If the analyst believes that the true boundary has high frequency fluctuations smoothed out by coarse sampling, the length scale should match the frequency of those fluctuations. Alternatively, if you are analysing the uncertainty from biases of a given imaging modality, such as thresholding or volume-reducing smoothing, the length scale could be matched to the size of the image domain. Due to the cheapness of generating different geometries, length scales can be determined through trial and error until the smoothness of the alternative geometries match analyst expectations.

\subsection{Surface and volume meshing}
Surface meshes are a fundamental building block in many spatial simulations. They are a flexible format for describing the geometry of complex objects and have a number of favourable properties. Generally, their structure consists of points in space, nodes, connected by lines in such a way that tessellating triangles or rectangles, facets, are formed to create a surface. These types of meshes describe the surface of an object and are used in almost every step of many BVP pipelines that involve geometries derived from images.

The flexible format of a surface mesh makes them resilient to small changes in its shape. The mesh is defined only by the locations of each node and the connectivity between them; therefore, each individual node can be moved slightly and the facets adapt accordingly. Additionally, each facet is a polygon with an easily calculable normal to its surface. To find the normal vector at any node in a mesh, the normal of each facet defined by that node is averaged. These properties make geometric morphing of surface meshes a very appealing proposition.

However, popular BVP solvers, such as the finite element method, cannot compute a solution using a surface mesh alone; they require a volumetric mesh to discretise the simulation domain into tessellating elements. These meshes are derived from a surface mesh using sophisticated mesh generation software but their space-filling structure means they cannot be manipulated as easily as their surface counterpart. Moving nodes in the volume mesh directly may risk inverting elements or otherwise invalidating the mesh and its numerical characteristics.

Previous node perturbation studies eliminate this risk by scaling the perturbation size by the cell size, ensuring the former never exceeds the latter \cite{Abdulle2021}. Unfortunately, the scale of the perturbation in the random boundary method is set by external, physical sources and as such, the perturbations cannot be guaranteed to maintain the integrity of a sophisticated volumetric mesh. Therefore, the surface mesh must be perturbed and a volumetric mesh must be generated for each new surface individually. This can incur significant expense as 3D meshing is a complex and costly procedure. Techniques such as the immersed boundary methods could be investigated as an avenue for reducing this computational overhead.

\subsection{Algorithm}
The following algorithm gives a general guideline to construct alternative geometries of a surface mesh:

\itemsep-1em
\begin{enumerate}
  \item Take the set of nodes and their positions, $X = \left\{x_1, ..., x_n\right\}$, that describe the geometry's surface.
  \item Find the unit vector, $\vec{n}_i$, that is normal to the local surface at node $x_i$. Most mesh manipulation codes will offer a method to do this step, usually by taking an area-weighted average of the normal vector for each of the facets attached to a given node.
  \item \label{mean} Assign a mean deviation from the standard geometry, $\mu$. In the absence of any perceived systematic biases in the geometric description, set $\mu = 0$. Global artefacts such as volume-reducing smoothing can be accounted for here.
  \item \label{kernel} Assume a spatial covariance structure for the uncertainty in geometry and describe it using a kernel such as Equation \ref{kernel_eq}.
  \item Generate a random field over the geometry's domain based on the kernel from Step \ref{kernel} and the mean deviation from Step \ref{mean}. This can be done using a number of methods including eigenvalue decomposition or a truncated Karhunen-Lo\`eve expansion for larger models. Many software packages exist to help with this step.
  \item Sample from the random field at each node and generate a scalar, $w(x_i) \sim \mathcal{N} (\mu, C(X, X'))$, at each node to form the set $W = \left\{w_1, ..., w_n\right\}$.
  \item For each node, update its position, $\vec{x}_i$, by moving it along its normal vector with a magnitude defined by the random draw at that node $w(\vec{x}_i)$,
  \begin{equation}
    \vec{x}_i \leftarrow \vec{x}_i + w(\vec{x}_i)\vec{n}_i
  \end{equation}
  The updated set of coordinates now defines a new, random geometry.
\end{enumerate}

Once a statistical sample of possible geometries has been generated, the simulation's sensitivity to uncertain geometries can be found by following the standard meshing and solving procedure for the simulation. A Julia example applying this procedure for STL files can be found in this paper's accompanying \href{https://github.com/gerrygralton/random_boundaries}{software repository}.

\section{Numerical examples}\label{num_examples}
The random boundaries method is demonstrated below with examples from linear elasticity and fluid mechanics. These examples are designed to show the wide applicability to different types of biomedical simulations while also highlighting the uncertainties associated with different types of poorly described geometries.

\subsection{Linear elasticity with a patient-specific geometry}\label{lin_elastic}
\begin{figure}
  \captionsetup{width=\linewidth}
  \begin{center}
  \includegraphics[width=0.85\linewidth]{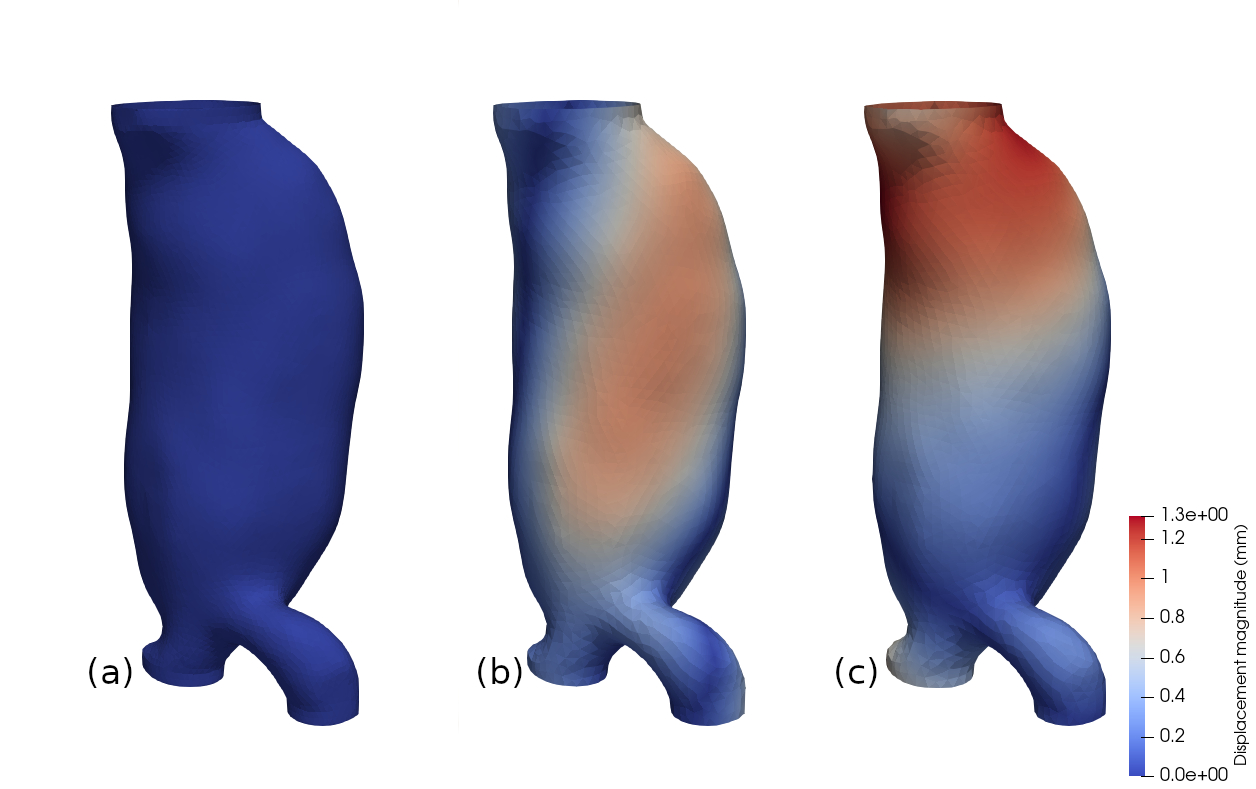}
  \end{center}
  \caption{The originally segmented geometry (a) compared to two random geometries (b and c) generated using a mean-zero Gaussian random field and a squared-exponential covariance function where $l=50$mm and $\sigma=0.624$mm (equivalent to one pixel's in-plane width). The visually subtle geometric variations from the segmented geometry are highlighted with the colour gradient.}
  \label{fig:aaacomp}
\end{figure}
Biomechanical modelling is currently being touted as a way to generate biomechanical indicators for the likelihood of a fatal rupture of an abdominal aortic aneurysms (AAA). Many different methods exist for calculating these biomechanical indicators but they share a common reliance on three-dimensional imaging to generate patient-specific geometries on which their simulations can be run \cite{Joldes2017,Polzer2013,VASCOPS,Bappoo2021}. The dependence on the geometries extracted from images begs the questions of how repeatable is this initial processing and what is the impact on these simulations' predictions.

BioPARR is a widely used simulation pipeline developed in 2017 to calculate stress in the wall of an abdominal aortic aneurysm (AAA) \cite{Joldes2017}. This software has been used here as a demonstration of a linear PDE's sensitivity to the uncertainty in an image-derived geometry. Popular rupture potential indexes such as \cite{Vande2006} which are intended to be an indicator of AAA disease severity take into account only the peak or 99th percentile maximum principal stress (MPS). \cite{Speelman2008} found that the 99th percentile MPS is a "more reproducible" and reliable parameter than the peak stress. As such, we will use the 99th percentile MPS in the aneurysm wall as our variable of interest and look to find its sensitivity to geometric perturbations. We apply random boundaries to a segmented image of a single patient's abdominal aortic aneurysm to find the uncertainty in the predicted stress.

A CT image with 0.624x0.624x1.0mm voxels of a patient's abdominal aortic aneurysm (AAA), a slice of which is shown in Figure \ref{fig:boundary_example}, was segmented and the geometry's surface was created using the method described in \cite{Joldes2017} and the tools contained within the 3D Slicer package \cite{Fedorov2012}. From this surface the method described in Section \ref{method} was applied and 100 alternative geometries were generated. A zero-mean Gaussian random field was used with a squared-exponential covariance kernel where $\sigma^2=0.624$mm (1 pixel width) and $l=50$mm. In this example the random field was generate using a truncated Karhunen-Lo\`eve expansion. Two of these artificially generated geometries are shown next to the original, segmented geometry in Figure \ref{fig:aaacomp}. The deviation magnitude, $\sigma$, was selected by analysing the boundary wall definition in the CT image and deciding that the standard deviation was approximately 1 pixel width. The length scale was selected to be similar to the length of the aneurysm in order to quantify systemic biases in the segmentation process, such as incorrect thresholding.

These modified geometries were then meshed into corresponding 3D volumetric meshes with a uniform 1.5mm wall thickness. Tetrahedral elements were used and the procedure was handled within the BioPARR framework using gmsh \cite{Geuzaine2009}. The default BioPARR parameters were used for the finite element analysis. The applied load was 0.016 Pa of uniformly applied internal pressure. The constitutive properties were chosen as a linear elastic wall with a Young's modulus of $10^5$ Pa and a Poisson ratio of $0.49$.

\begin{figure}
  \captionsetup{width=\linewidth}
  \begin{center}
  \includegraphics[width=0.85\linewidth]{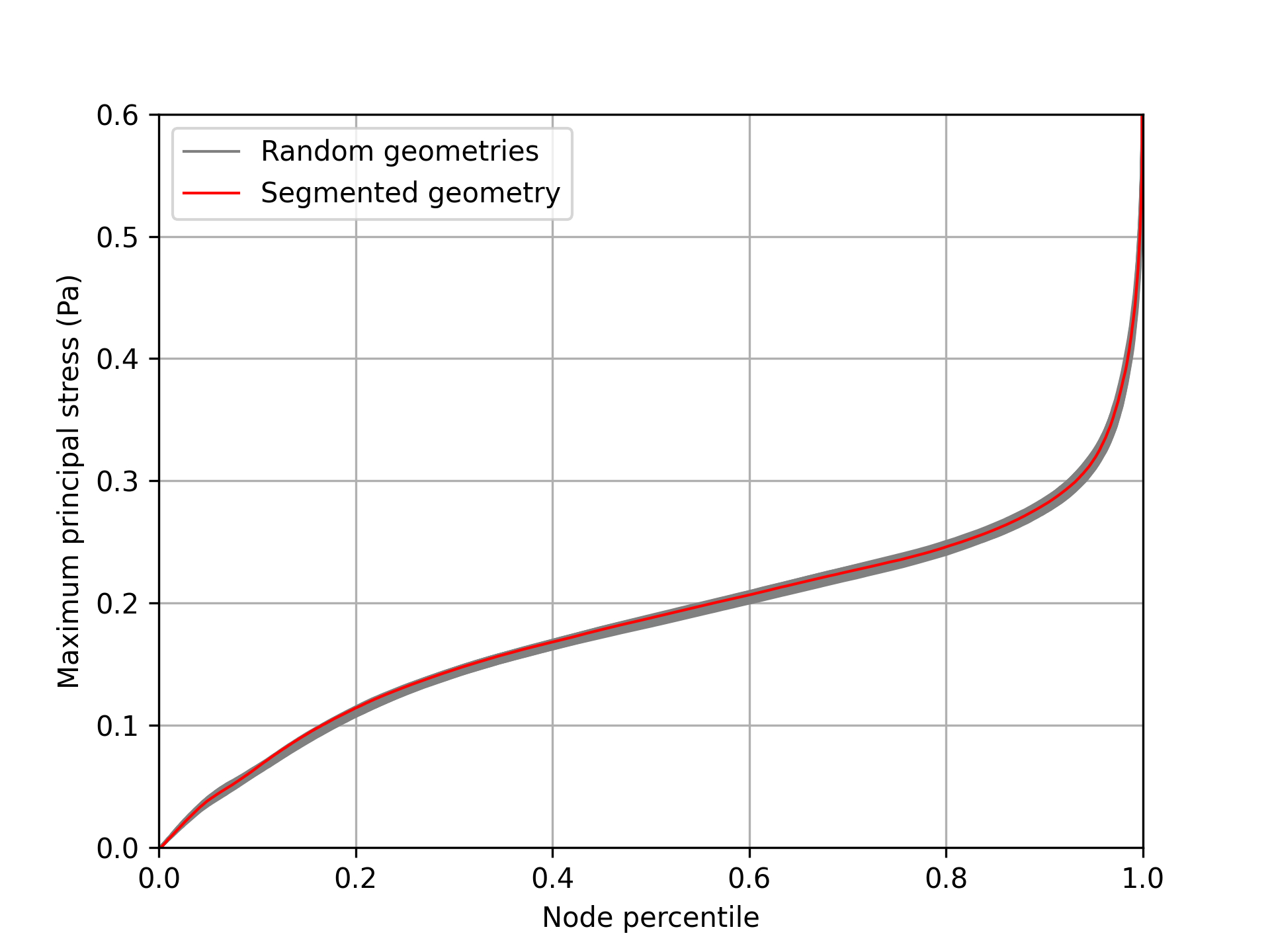}
  \end{center}
  \caption{Maximum principal stress in a segmented AAA geometry and 100 randomly generated alternative geometries. }
  \label{fig:mps_dist}
\end{figure}

\begin{figure}
  \captionsetup{width=\linewidth}
  \begin{center}
  \includegraphics[width=0.85\linewidth]{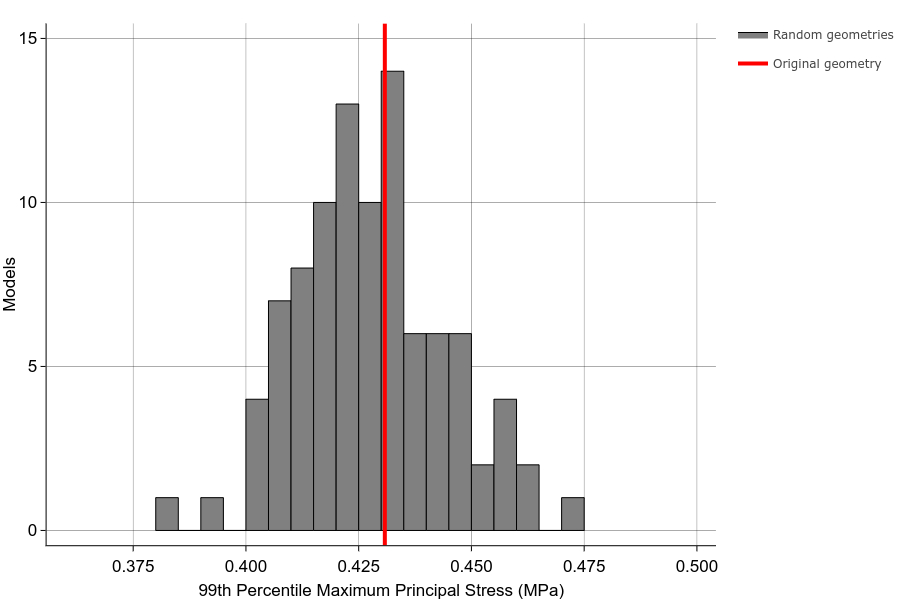}
  \end{center}
  \caption{Distribution of 99th percentile maximum principal stress in the segmented AAA geometry and randomly generated alternative geometries.}
  \label{fig:mps_99_hist}
\end{figure}
The resulting maximum principal stress (MPS) distributions are shown in Figure \ref{fig:mps_dist}. Looking at the 99th percentile of these distributions, shown in Figure \ref{fig:mps_99_hist}, it can be seen that there is a significant spread in 99th percentile MPS. The 95\% confidence interval on this resulting distribution is between 0.398-0.463 MPa with the segmented geometry returning a stress of 0.4308 MPa. It is worth noting that this uncertainty is significantly less dramatic than the study published alongside BioPARR where 3 different analysts, including a layperson, generated geometries resulting in up to 25\% differences in MPS \cite{Joldes2017}. A similar study in \cite{Hodge2022} found inter-user difference to generate up to 35\% difference in 99th percentile MPS. This helps reinforce our assumption that gross differences between analysts are unaccounted for when using the random boundaries methodology.

\subsection{Poiseuille flow in an idealised geometry}
\begin{figure}
  \captionsetup{width=\linewidth}
  \begin{center}
  \includegraphics[width=0.85\linewidth]{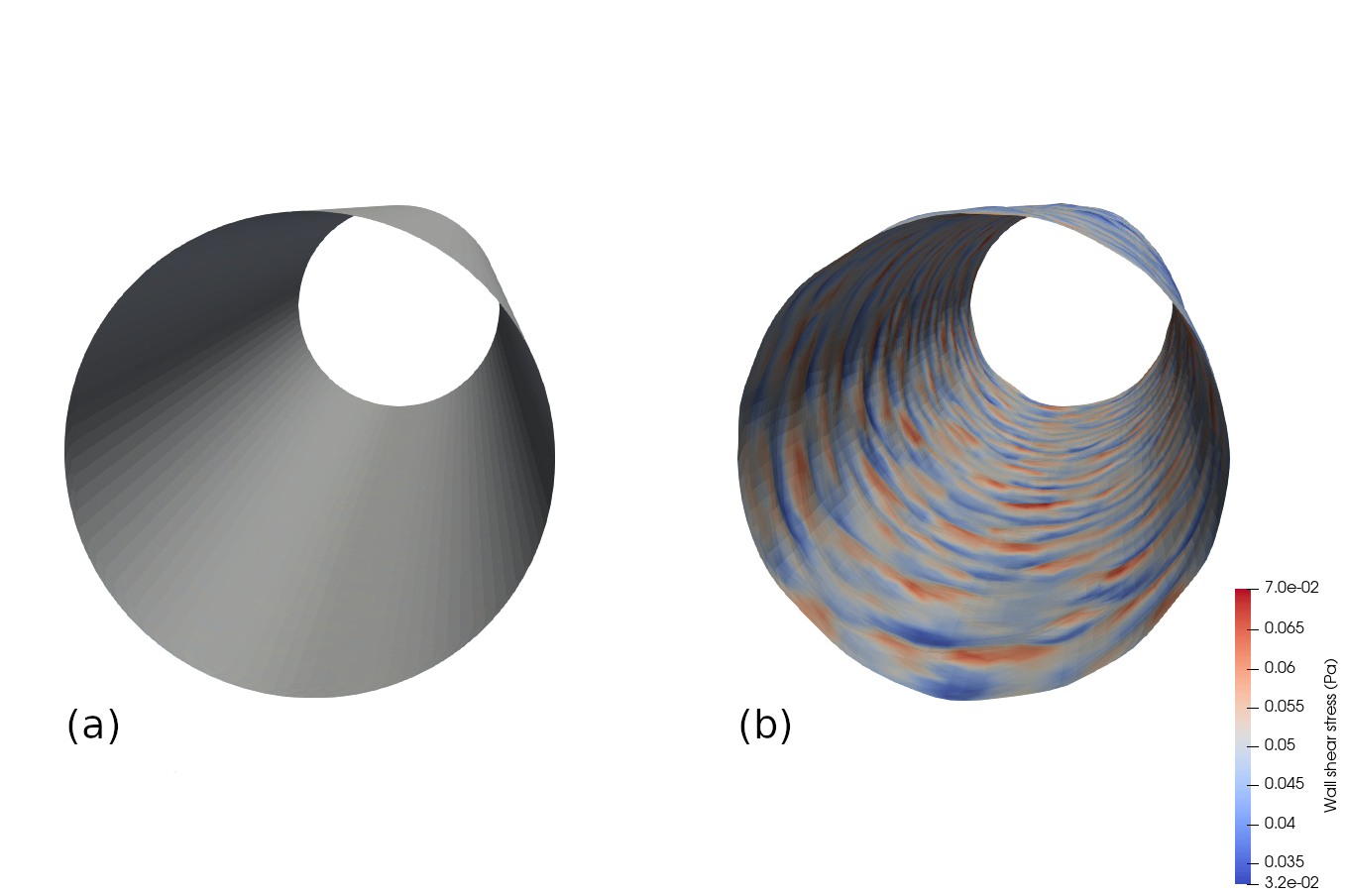}
  \end{center}
  \caption{Wall shear stress on a perfect cylinder (a) and a random alternative geometry (b). The analytic solution for wall shear stress for wall shear stress from Poiseuille flow in a perfect cylinder is 0.05 Pa.}
  \label{fig:cyls}
\end{figure}
\begin{figure}
  \captionsetup{width=\linewidth}
  \begin{center}
  \includegraphics[width=0.85\linewidth]{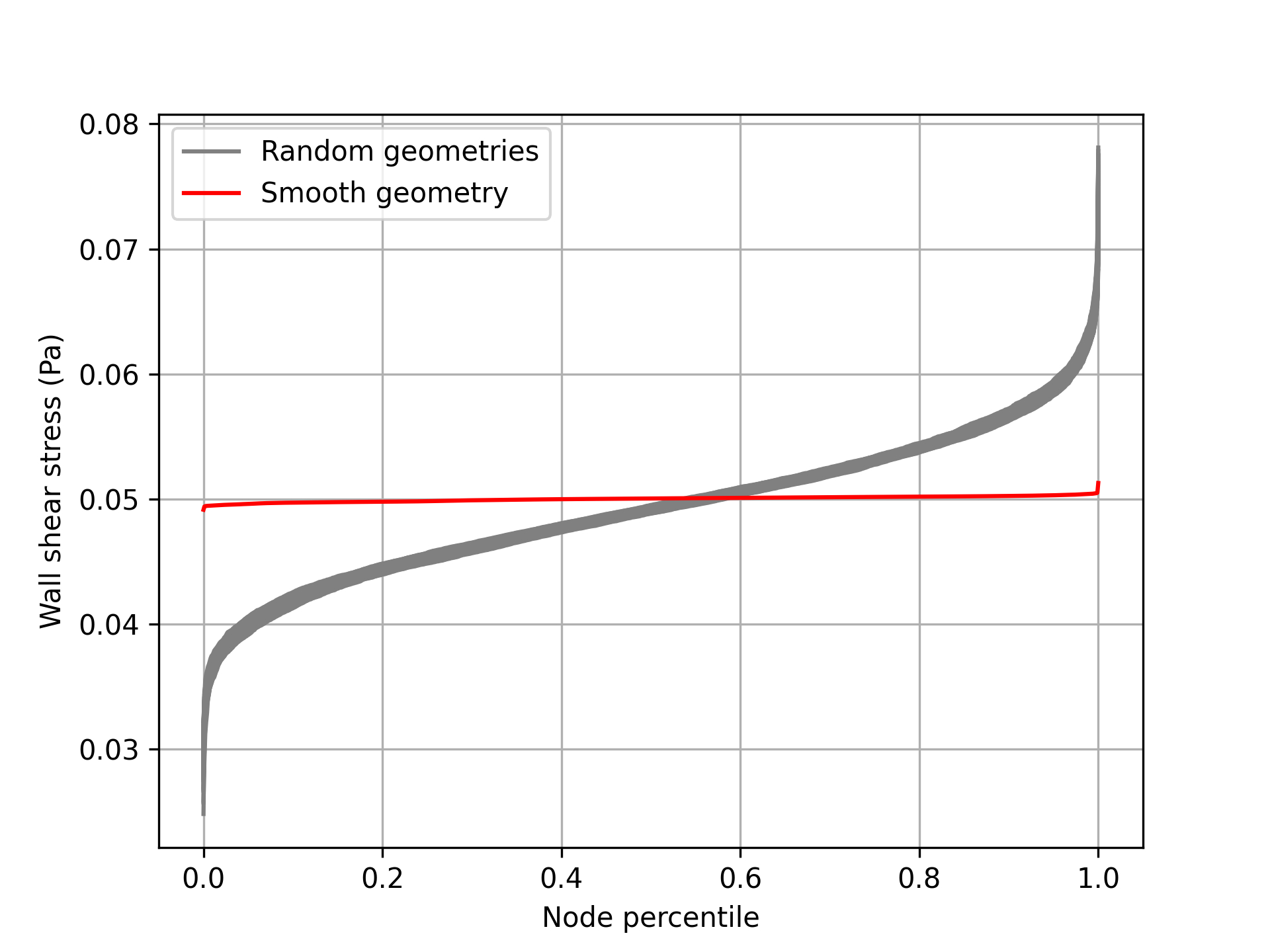}
  \end{center}
  \caption{Nodal wall shear stress values on a perfect cylinder and 100 random alternative geometries. The analytic solution for wall shear stress from Poiseuille flow in a perfect cylinder is 0.05 Pa.}
  \label{fig:cyl_histogram}
\end{figure}
This example is from a benchmark problem for a large body of literature that uses wall shear stress as a clinical indicator of coronary disease \cite{Morris2016,Bappoo2021}. Typically, these works assume that an unusually high or low wall shear stress value in a given blood vessel is associated with material degradation, growth or rupture \cite{Slager2005,Sforza2012,Dolan2012,Samady2011}. Wall shear stress is found by simulating viscous flow through a pipe with a geometry derived from a patient-specific, computed-tomography angiogram. This imaging modality generates a high-contrast but low-resolution, two-dimensional projection of coronary blood vessels. Patient specific computational models can then generated from these images by extracting the coronary vessel's two-dimensional centerline, sampling the vessel width at given intervals along this path and connecting these locations to form a channel or tube. This technique makes an inherent assumption that the wall is locally smooth. Other segmentation procedures also exist but have similar implications for local smoothness.

The impact of this local smoothness assumption has been demonstrated here using an ideal, straight cylinder with a radius of 5mm and a length of 50mm. A Poiseuille flow was computed using FeNICS's finite element method (FEM) solver \cite{Logg2012} with P2P1 elements, a dynamic fluid viscosity of $0.00345$ Ns/m\textsuperscript{2} and a density of $1050$ kg/m\textsuperscript{3}.  Pressure boundary conditions were applied such that $p_{inlet} = 1$ Pa and $p_{outlet} = 0$ Pa while a non-slip velocity boundary was established along the cylinder walls. The time step used was 0.015 seconds and velocity field was deemed to have reached steady-state when accelerations were found to be less than $10^{-6}$ m/s\textsuperscript{2}. The simulation took 1230 seconds as a parallel computation on a 16 thread, 2.3GHz processor with 32GB of memory. Once an equilibrium flow had been established, post-processing using FeNICs calculated the shear stress on the wall.

After simulating this ideal scenario, 100 alternative cylinders were generated using an eigen-value decomposition method with a mean deviation of zero and a squared-exponential covariance where $\sigma = 0.05$mm and $l=1$mm. These parameters, especially $\sigma$, are two orders of magnitude below the resolution of a standard angiogram and are intended to show the potential ramifications of small fluctuations in the surface of the blood vessel that would be impossible to perceive with this imaging modality.

The results in Figure \ref{fig:cyls} show that the perfect cylinder has an almost constant wall stress over its surface. This is in agreement with the analytic solution for Poiseuille flow in this cylinder which gives a value for wall shear stress of 0.05Pa. Figure \ref{fig:cyls} shows that even with a very small perturbations in the wall's surface, the resulting shear stress distribution becomes significantly larger, with some nodes experiencing stress that is both 40\% higher and lower than when modelled with an ideal cylinder. The distribution of nodal wall shear stress values on each model are shown in Figure \ref{fig:cyl_histogram} to highlight the dramatic shift from the smooth solution when small geometric deviations are introduced.

This finding is of practical and clinical importance. A significant amount of research and commercial interest has been dedicated to these cardio-vascular fluid simulations and the above results give reason to investigate their sensitivity to geometry. The proposed random boundaries technique provides a reliable, repeatable way to parameterise this uncertainty.

\section{Limitations and further work}
The proposed technique for quantifying segmentation uncertainty in solutions to boundary-value problems has a number of limitations worthy of discussion. Chiefly, random boundaries uses a Monte-Carlo simulation method which takes significant computational resources to generate a distribution. Other methods, relying on more efficient sampling, may be more appropriate and it would be of enormous benefit to integrate this in any further applications.

Random boundaries makes an explicit assumption that the uncertainty in a given boundary's location can be modelled with a Gaussian distribution; this does not necessarily reflect the reality of image-based segmentation uncertainties. Further extensions of this technique could focus on verifying this assumption against other possible distributions.

Additionally, it may be of benefit to complete a more thorough study into the comparability of the random boundaries technique to those uncertainty quantification methods already present in the literature. A small, initial example has been included in Section \ref{lin_elastic} but could be further extended.

\section{Conclusions}
This paper presents a powerful new method for parameterising geometric uncertainty and quantifying a boundary value, partial differential equation's response. It is shown that various types of simulations can be sensitive to geometric uncertainty and that this source of uncertainty should be regarded with as much caution as any other BVP input. We hope that this goes some way towards encouraging practitioners, especially in bioengineering fields, to consider the effects of geometric uncertainty. In the context of image segmentation, it is widely acknowledged that different analysts will generate notably different geometries but the effects of this are opaque. By framing this problem as an image analysis issue and quantifying our geometric uncertainty using the resolution limitations of the image itself we can standardise and compare the sensitivity of dependent algorithms with much greater clarity.

\subsubsection{Acknowledgements} The first author is a recipient of the Research Training Program (RTP) scholarship and a co-funded University Postgraduate Award scholarship at the University of Western Australia.\\
Patient data was obtained with approval from Human Research Ethics and Governance at South Metropolitan Health Service (HREC-SMHS) (approval code RGS3501) and The University of Western Australia Human Research Ethics Committee (approval code RA/4/20/5913).\\
We gratefully acknowledge support of the Australian Government through the National Health and Medical Research Council NHMRC (Project Grant No. APP1162030 and Ideas Grant No. APP2001689) and Australian Research Council ARC (Discovery Project DP160100714).\\
We also thank Fiona Stanley Hospital, Perth, Western Australia for providing the patient CT images.

%
%
%
\bibliographystyle{splncs04}
\bibliography{random_boundaries}

\end{document}